\documentclass{article}
\usepackage{times,graphicx,a4wide,natbib}

\begin{document}
	
\title{The Effect of Probe Dynamics on Galactic Exploration Timescales}
\author{Duncan H. Forgan$^1$, Semeli Papadogiannakis$^1$ and Thomas Kitching$^1$}
\maketitle

\noindent $^1$Scottish Universities Physics Alliance (SUPA), Institute for Astronomy, University of Edinburgh, Blackford Hill, Edinburgh, EH9 3HJ, UK \\

\noindent \textbf{Direct Correspondence to:} \\
\textbf{Post:} Dr Duncan Forgan \\
East Tower, Institute for Astronomy \\
University of Edinburgh \\
Blackford Hill \\
EH9 3HJ, UK\\
\textbf{Email:} dhf@roe.ac.uk

\newpage

\begin{abstract}

The travel time required for one civilisation to explore the Milky Way
using probes is a crucial component of Fermi's Paradox.  Previous
attempts to estimate this travel time have assumed that the probe's
motion is simple, moving at a constant maximum velocity, with powered
flight producing the necessary change in velocity required at each
star to complete its chosen trajectory.  This approach ignores lessons
learned from interplanetary exploration, where orbital slingshot
maneouvres can provide significant velocity boosts at little to no
fuel cost.  It is plausible that any attempt to explore the Galaxy would
utilise such economising techniques, despite there being an upper limit
to these velocity boosts, related to the escape velocity of the object
being used to provide the slingshot.

In order to investigate the effects of these techniques, we present
multiple realisations of single probes exploring a small patch of the
Milky Way. We investigate 3 separate scenarios, studying the slingshot
effect on trajectories defined by simple heuristics.  These scenarios
are: i) standard powered flight to the nearest unvisited star without using
slingshot techniques; ii) flight to the nearest unvisited star using
slingshot techniques, and iii) flight to the next unvisited star which
provides the maximal velocity boost under a slingshot trajectory.

We find that adding slingshot velocity boosts can decrease the travel
time by up to two orders of magnitude over simple powered flight. In
the third case, selecting a route which maximises velocity boosts also
reduces the travel time relative to powered flight, but by a much
reduced factor.  From these simulations, we suggest that adding
realistic probe trajectories tends to strengthen Fermi's Paradox.

Keywords: Fermi Paradox, SETI, interstellar exploration, probe dynamics
\end{abstract}

\section{Introduction}

\noindent Fermi's Paradox remains an important cornerstone of modern
thinking on extraterrestrial intelligence (ETI).  It taxes most, if not all
attempts to formulate an optimistic perspective on the frequency of
alien civilisations in the Galaxy both in space and time.

Detailed reviews of the Paradox can be found in \citet{BrinG.D.1983},
\citet{Webb2002} and \citet{Cirkovic2009}.  The Paradox rests on the
current absence of ETI in the Solar System (what \citealt{Hart1975} refers to as
``Fact A'').  This absence runs counter to estimated timescales for
intelligent species to colonise of the Galaxy - what
\citet{Cirkovic2009} (and references within) refers to as the
Fermi-Hart timescale (see e.g. \citealt{Hart1975,Tipler1980}):

\begin{equation} t_{FH} = 10^6 - 10^8 \,\mathrm{yr}\end{equation}

\noindent This is compounded by the fact that the age of the Earth is
at least an order of magnitude higher, and the median age of terrestrial
exoplanets is estimated to be a further 1 Gyr older than the Earth
\citep{Line_planets}.  It appears the inexorable conclusion one must
draw is that ETIs do not exist, otherwise we would have detected their
presence in the Solar System.  

It is common for attempts to resolve the Paradox to speculate on the
motivation or sociological make-up of ETIs - for example, one solution
suggests that the Earth has attained a protected status and must not
be disturbed, often known as the Zoo Hypothesis
(e.g. \citealt{Ball1973}).  While flaws can be exposed in these types
of hypothesis (see e.g. \citealt{Forgan2011b}), there are many
solutions that are simply unfalsifiable, and while they cannot be
ruled out, they cannot be currently considered as scientifically
meritorious.  Until conclusive data is compiled on something as
esoteric as extraterrestrial sociology, it is more worthwhile to focus
on potential physical constraints for extraterrestrial contact.

Weaker formulations of the Paradox merely rest on ETIs practising
interstellar communication rather than interstellar colonisation (e.g.
\citealt{Scheffer1994})- stronger formulations use self-replicating Von
Neumann probes to explore the galaxy at an exponential rate.  It is
this process of self-replication (or colonies carrying out subsequent
autonomous colonisation) that allows $t_{FH}$ to be small enough for
the Paradox to be robust.  While there have been many arguments for
and against the use of self-replicating probes
(e.g. \citealt{Tipler1980, Sagan1983,Chyba2005,Wiley2011}), we wish to
focus instead on a more fundamental aspect of probe exploration that
has not been addressed fully.

The Paradox leans heavily on the dynamics of interstellar flight, and
the motivations of the extraterrestrial intelligences (or ETIs) that
drive the exploration.  \citet{Sagan1963} expounds a framework for a
cadre of civilisations visiting star systems using relativistic
interstellar flight.  Under these assumptions, the visiting rate for
main sequence stars could be as high as once per ten thousand years
(although it makes relatively optimistic assumptions about the number
of civilisations forming the cadre).  The associated problems of
population growth and carrying capacity are also important drivers for
continual exploration, as was explored by \citet{Newman1981} using
nonlinear diffusion equations.  The stipulation that ETIs practise
zero population growth can alter $t_{FH}$ by several orders of
magnitude.

At a more fundamental level, \citet{Bjørk2007} investigated probe
exploration in a schematic model of the Galaxy, with an exponentially
declining stellar surface density distribution.  Each host probe
visits a subset of the Galactic stellar population, containing 40,000
stars.  The host then releases 8 sub-probes, which explore this
subset, and then dock with the host before travelling to a new
Galactic sector.  Each probe has a constant velocity of $0.1c$, and
travels to its nearest unvisited star under powered flight.  There is
no attempt to optimise the trajectory of the sub-probes beyond this
simple heuristic.  Optimising its trajectory is an instance of the
``travelling salesman'' problem \citep{Golden1988,Toth2001}, an
NP-hard problem which is computationally prohibitive to solve for
large node numbers, as is the case in interstellar exploration.

\citet{Cotta2009} extended this work by using algorithms
to improve the trajectory of the probes, reducing the tour length by
around 10\%.  They subsequently carried out a probabilistic analysis
assuming many ETIs were carrying out exploration in this fashion.
This allowed them to estimate the maximum number of ETIs exploring the
Galaxy ($N_{ETI}$) that could do so while ensuring contact with Earth remained
statistically unlikely.  In the parameter space explored, they
estimate that the upper limit is approximately $N_{ETI} < 10^3$.

What is common to all these simulations of probe exploration is that
they ignore or neglect lessons learned from Mankind's unmanned
exploration of the Solar System.  Simple powered flight is an
inefficient means of travel, especially for probes using chemical or
nuclear propulsion methods.  Using slingshots inside gravitational
potential wells allows the probe to produce relatively large $\Delta
v$ and alter its trajectory without expending fuel, and potentially
boosting its speed relative to the rest frame of its starting
position.  This has been utilised successfully by Mankind during its
exploration of the Solar System, both in the ecliptic plane through
the triumphal tour of the Voyager probes (see e.g. \citealt{Gurzadyan1996}), and
even out of the ecliptic during the Ulysses mission \citep{Smith1991}.

In principle, this behaviour can be scaled up to the Galactic level,
where the potential well of stars can now be used to provide $\Delta
v$ and increase the probe speed relative to the Galactic rest frame
\citep{Surdin1986}.  While this requires probes to be able to navigate
extremely accurately, this would not appear to be an insurmountable
obstacle provided the probe possessed sufficient autonomous computing
power, and retained enough fuel for judicious course corrections.

In this paper, we investigate the effect of individual probe dynamics
on the visitation timescale of a population of stars.  We run a series
of Monte Carlo Realisations (MCRs), where in each realisation a single
probe traverses a path through a population of stars.  The stellar
population has a number density and velocity field akin to that of the
Solar Neighbourhood.  We run three separate sets of realisations,
focusing on three basic scenarios:

\begin{enumerate}
\item A single probe, visiting stars under powered flight, with each
  leg of the route determined by finding the nearest unvisited
  neighbour (which we label \textbf{powered}),
\item As 1., except utilising slingshot trajectories to boost the
  velocity of the probe (which we label \textbf{slingshot}),
\item A single probe which selects the next star to travel to such
  that the velocity boost derived from a slingshot is maximal, which
  depends on the destination star's velocity relative to the current
  star (which we label \textbf{maxspeed}).
\end{enumerate}

\noindent In section \ref{sec:method}, we describe the mechanics of
the slingshot maneouvres employed in this paper, as well as the setup
of the simulations; section \ref{sec:results} describes the results of
the three scenarios, and in sections \ref{sec:discussion} and
\ref{sec:conclusions} we discuss and conclude the work.

\section{Method}\label{sec:method}

\subsection{The Dynamics of Slingshot Trajectories}\label{dynamics}

A slingshot trajectory uses the momentum of the star it passes to gain
or lose velocity, depending on the incident angle of the probe's
approach.  The probe therefore does not need to expend additional
energy that would otherwise be required to complete a similar
trajectory under powered flight. We briefly describe the mathematics
of slingshot trajectories here: for more detail the reader
is referred to \citet{Gurzadyan1996} (in particular to Chapter XIII,
section 4, equations 29 to 31).

\begin{figure*} \label{fig:nextstar}
\begin{center}$
\begin{array}{cc}
\includegraphics[scale = 0.4]{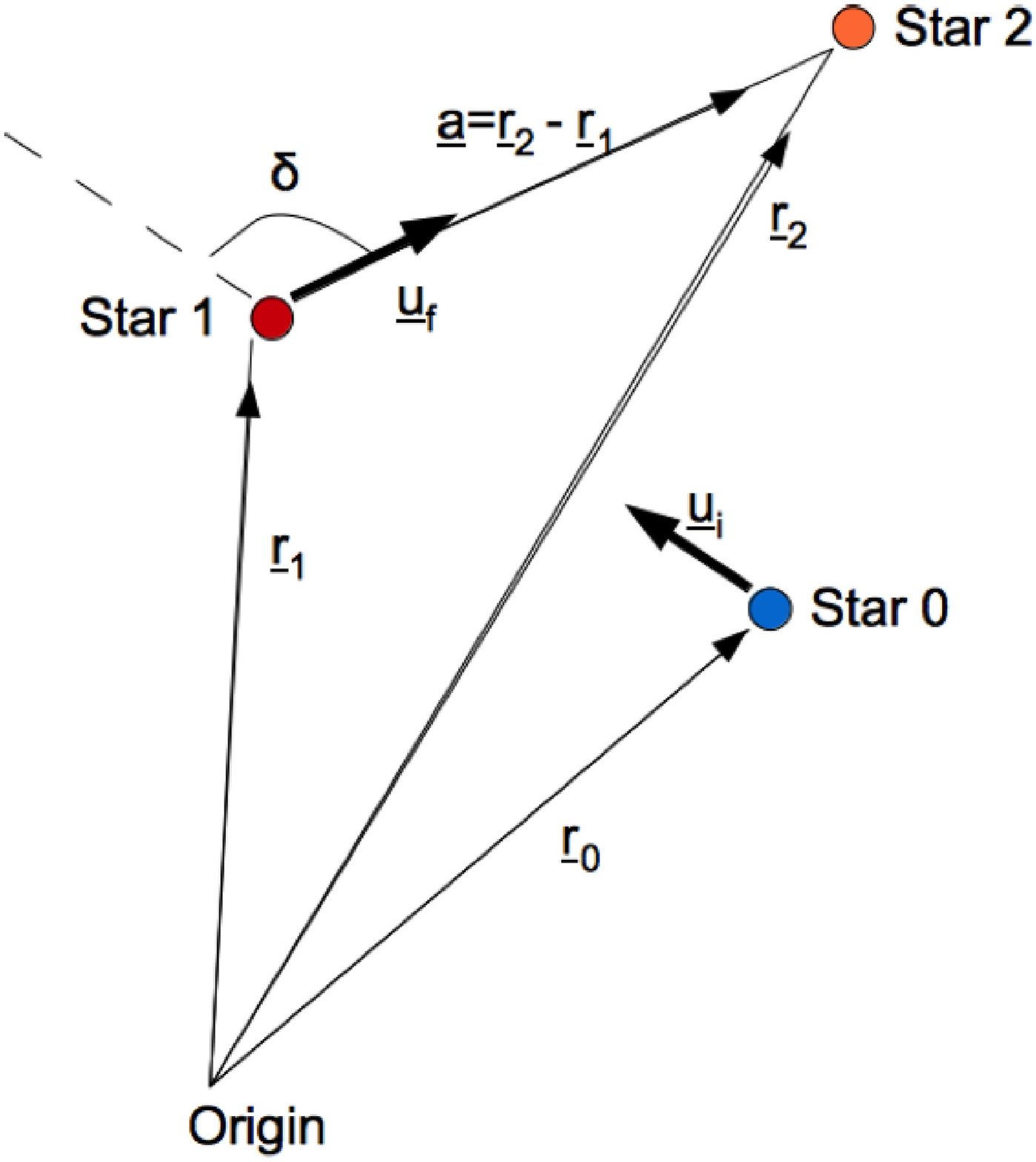} &
\includegraphics[scale = 0.6]{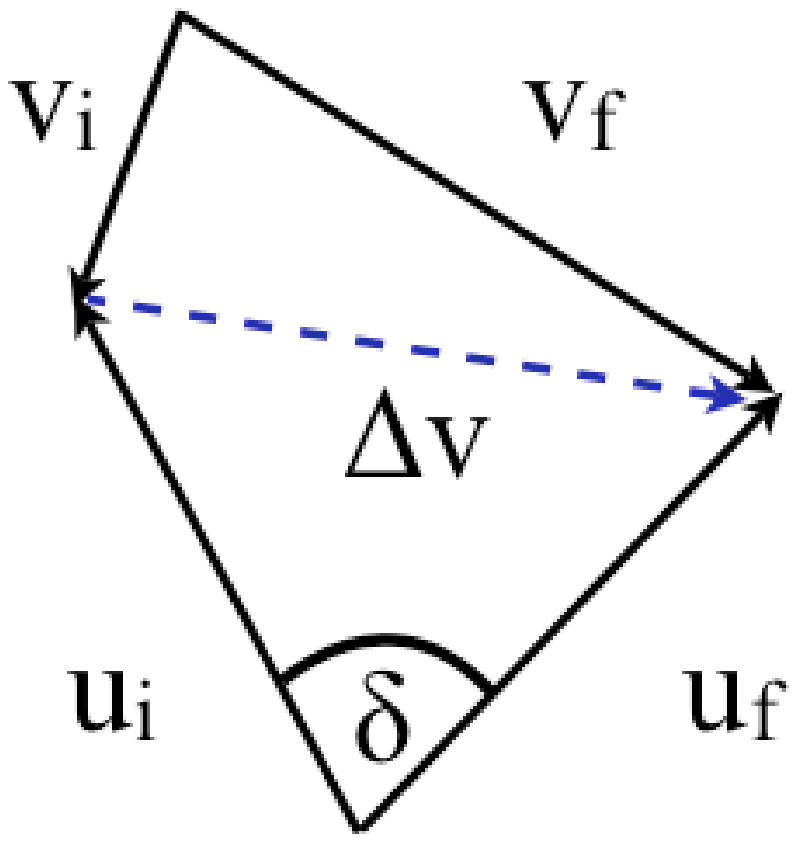} \\
\end{array}$
\caption {The vector diagram above shows he path of the probe from the
  star’s reference frame, changing its direction by an angle $\delta$
  but not the magnitude of its velocity.}
\end{center}
\end{figure*} 

The left panel of Figure \ref{fig:nextstar} describes one stage of the
probe's trajectory when slingshots are used.  The probe begins at Star
0, accelerating to velocity $\mathbf{u}_i$ (measured in the frame
where Star 1 is at rest).  For the probe to proceed to star 2 after
the slingshot, its velocity must assume the vector $\mathbf{u}_f$,
again in Star 1's frame.  The angle between $\mathbf{u}_i$ and
$\mathbf{u}_f$ is $\delta$.  In this manoeuvre, the probe follows a
hyperbolic trajectory, and

   \begin{equation} |\mathbf{u}_i| = |\mathbf{u}_f|. \end{equation} 

However, we are interested in the probe's velocity in the ``lab''
frame, i.e. a frame of reference where the Galactic Centre is at rest.  The
right panel of Figure \ref{fig:nextstar} shows the vector diagram
connecting the initial and final velocities in both frames.  As a
result of adding the star's motion to $\mathbf{u}_i$ and
$\mathbf{u}_f$ to create the Galactic frame velocities $\mathbf{v}_i$
and $\mathbf{v}_f$, we can see that there is indeed an increase in the
probe's speed, $\Delta v$, which we can deduce simply:

\begin{equation} \label{eq:target}
\Delta v = 2 |\mathbf{u}_i| \sin \left( \frac{\delta}{2}\right)
\label{eq:deltav}.\end{equation}

\noindent The change in the probe's momentum in this frame is balanced
by a change in the star's momentum relative to the Galactic Centre.
The fractional decrease in momentum the star experiences is so small
that we can effectively regard it as negligible.  The angle $\delta$
is related to the inward velocity as follows:

\begin{equation} \tan \left(\frac{\delta}{2}\right) = \frac{GM_*}{r_c
    u_i} \label{eq:maxdelta}\end{equation}

\noindent where $r_c$ is the distance of closest approach to the
star.  This places an upper limit on the value of $\delta$ and
consequently $\Delta v$ (see Discussion).

We assume the probe has a maximum velocity it can attain under its own
power.  This defines the probe's initial velocity as it travels from
Star 0 to Star 1.  Henceforth, it can travel with increasing speed as
it undergoes a series of slingshot maneouvres.  The magnitude of the
boost achieved by the slingshot maneouvre is increased if the star's
own velocity runs parallel to the probe's trajectory.  Therefore, it
is possible that a probe could choose a course based on the spatial
velocities of stars relative to each other, such that it carries out
slingshots with maximal $\Delta v$.

\subsection{Simulating Probe motion}

\noindent We carry out three simulation scenarios, all in a small
patch of the Galaxy containing one million stars, at a uniform density
of 1 star per cubic parsec.  The stars were set in a shearing sheet
configuration to emulate the rotation curve of the Milky Way.  For
convenience, the stars were given velocity vectors for the slingshot
calculation, but fixed in position.  For each scenario, 30
realisations were carried out - this number represents a balance
between reducing computational expense and maintaining a sufficiently
small standard error arising from random uncertainties. As we will see
in later sections, 30 realisations is sufficient to produce total
travel times at accuracies of a few percent.  In each
realisation, the probe is allocated a different starting star.

The three scenarios are:

\begin{enumerate}
\item Powered flight to the nearest neighbour (\textbf{powered}).  The probe
  travels from the starting star to the closest neighbour at its
  maximum powered velocity. The $\Delta v$ is therefore fixed by the
  repeated deceleration and acceleration the probe makes at every
  stage of the trip.

\item Slingshot assisted flight to the nearest neighbour (\textbf{slingshot}). Here
  the path is identical to that taken when using powered flight due to
  the same method of choosing the next star.  However, the probe need
  only expend $\Delta v$ to accelerate to maximum velocity once, and
  does not need to decelerate, instead using the slingshot manoeuvre
  to repeatedly boost its maximum velocity.  We assume that the
  $\delta v$ expended by the probe to make course corrections to adopt
  a slingshot trajectory is negligible.

\item Slingshot assisted flight seeking the maximum velocity boost
  (\textbf{maxspeed}).  This selects a different path entirely to the
  other two scenarios, seeking instead the course such that the
  relative velocity vector between the current and destination stars
  is large and negative, i.e. the destination star is moving toward
  the current star.  This will give a larger velocity boost, but will
  in general require a longer path length to achieve it.
\end{enumerate}

\noindent We select a relatively low maximum velocity of $3\times
10^{-5} c$, where $c$ is the speed of light \emph{in vacuo}. This is
comparable to the maximum velocities obtained by unmanned terrestrial
probes such as the Voyager
probes\footnote{http://voyager.jpl.nasa.gov/mission/weekly-reports/index.htm}.
Admittedly, the Voyager probes achieved these speeds thanks to
slingshot trajectories, so the top speed of human technology under
purely powered flight is unclear.  To some extent, the maximum powered
speed of the probe is less important - increasing or decreasing this
maximum will simply affect the absolute values of the
resulting travel times in a similar fashion.  What is more important
is the \emph{relative} effect of changing the propulsion method and/or
the trajectory.

\section{Results}\label{sec:results}

\noindent In Table \ref{tab:params} we summarise the three scenarios
in terms of the total travel time to traverse all stars in the
simulation, averaged over all 30 realisations.  As the probe velocity
we have selected is particularly low, the total travel times are quite
large.  In fact, two scenarios (powered and maxspeed) have travel times
longer than the current age of the Universe ($1.37 \times 10^{10}$
yr), despite the standard error on the mean (which is generally a few
percent of the mean).  Therefore, it is unlikely that probes
travelling at maximum velocities for current human technology can
explore even a small fraction of their local neighbourhood in a timely
fashion.  We will investigate the reasons for this in more detail in
the following sections.

On the other hand, the slingshot scenario has a significantly shorter travel
time than the others.  Of course, this travel time is only for $10^6$
stars, and the Galaxy itself contains around $10^{11}$ stars, so it
still remains unlikely that probes of this type could explore the
entire Galaxy before Mankind began to construct devices that could
detect them.  For the Fermi Paradox to hold, the initial probe
velocity would need to be much larger, a fact that has long been obvious
\citep{Sagan1963}.

\begin{table}
\centering
  \caption{Summary of the three scenarios investigated in this work.\label{tab:params}}
  \begin{tabular}{c || cc}
  \hline
  \hline
   Simulation & Average Total Travel Time (yr) & Standard Error (yr) \\  
 \hline
  powered & $4.51 \times 10^{10}$ & $2.82 \times 10^{7}$  \\
  slingshot & $3.99 \times 10^{8}$ & $3.51 \times 10^{5} $  \\  
  maxspeed & $1.99 \times 10^{10}$ & $1.23 \times 10^{7} $\\
 \hline
  \hline
\end{tabular}
\end{table}

\subsection{Powered Flight Only, Nearest Neighbour (powered)}

\begin{figure}
\begin{center}
\includegraphics[scale = 0.5]{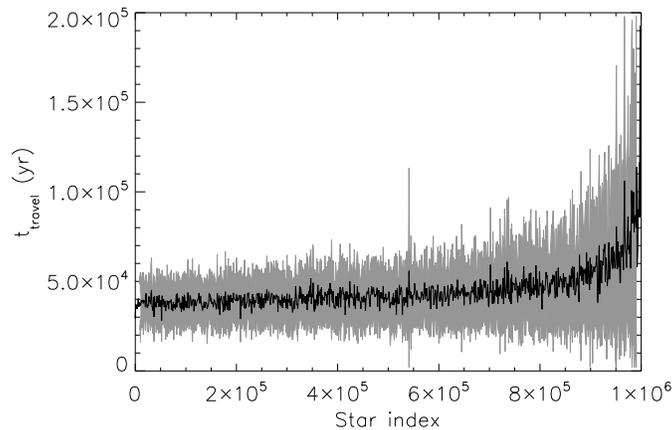} 
\caption{The mean travel time between each star for probes in the
  \emph{powered} simulations, averaged over 30 realisations.  The grey shaded
  area represents the standard error on the mean. }\label{fig:powered}
\end{center}
\end{figure}

\noindent Figure \ref{fig:powered} shows the travel time between each
star in the powered case.  Here we average over 30 realisations, with the
mean drawn in black and the standard error on the mean plotted in
grey.  Note that as the density of stars is roughly constant
throughout the box, and the $\Delta v$ achievable by the probe is
fixed by its own engines, the travel time between nearest neighbours is also
reasonably constant.  As the number of unvisited stars drops below a
hundred thousand or so, the probe must travel longer distances to find
an unvisited star, and hence the travel time increases towards the end
of the simulation.

\subsection{Slingshot Trajectory, Nearest Neighbour (slingshot)}

\begin{figure*}
\begin{center}$
\begin{array}{cc}
\includegraphics[scale = 0.45]{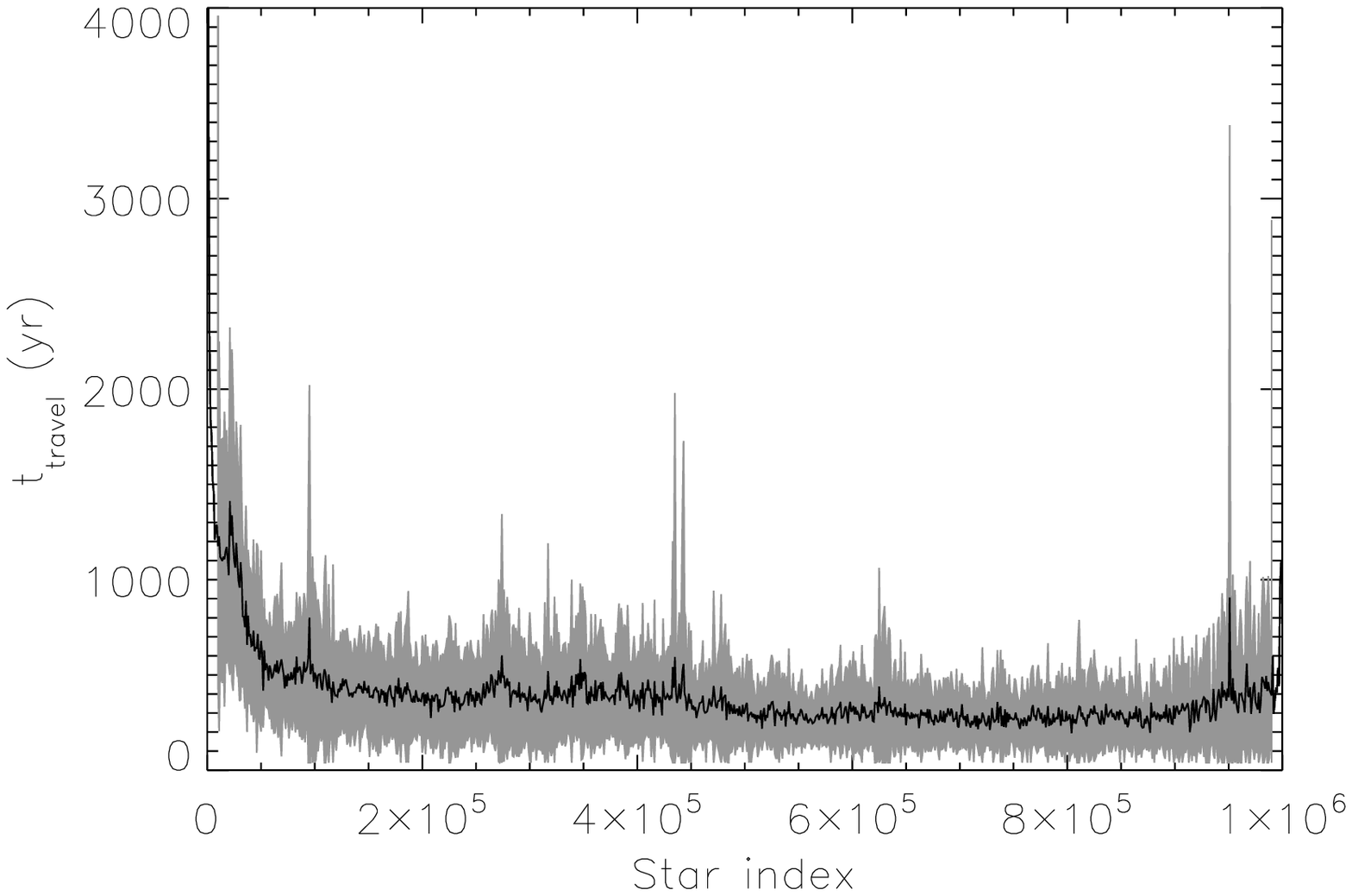} &
\includegraphics[scale = 0.45]{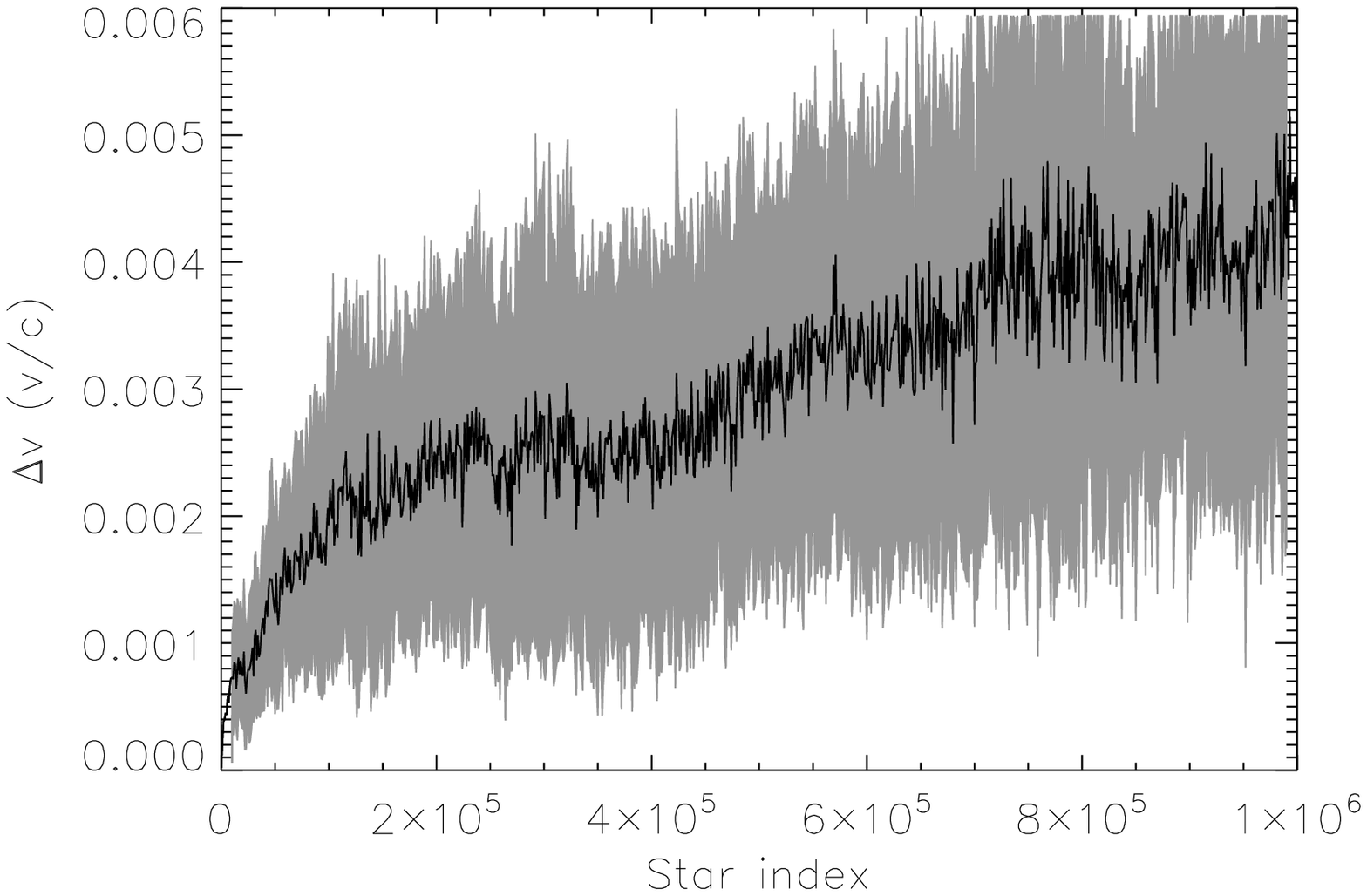} \\
\end{array}$
\caption{\textbf{Left}:The mean travel time between each star for
  probes in the \emph{slingshot} simulations, averaged over 30
  realisations.  The grey shaded area represents the standard error on
  the mean.  \textbf{Right}: the mean $\Delta v$ generated as a result
  of each slingshot maneouvre, with standard error on the mean shaded
  grey.} \label{fig:slingshot}
\end{center}
\end{figure*}

\noindent If we now allow the probe to make slingshot maneouvres to
its nearest neighbour (slingshot), then we can see that the probe's
behaviour changes significantly (Figure \ref{fig:slingshot}).  It
follows the same course as before, but it can now boost its velocity
at every stage of the journey (right panel), significantly reducing
its travel time (left panel).  As a result, it has increased its
velocity by almost a factor of 100 throughout the simulation (a fact
borne out by its total travel time being approximately 100 times
smaller than the powered case).  Even towards the end of its path,
where the unvisited nearby stars reside at larger distances, the
increased speed as a result of the slingshots allow the probe to cover
these larger distances in a short time span.  Note the large standard
error on the mean - $\Delta v$ is now a sensitive function of the
probe's path (more specifically, the angle $\delta$ in Figure
\ref{fig:nextstar}).  As the probe begins from a different starting
star in each realisation, the probe's path is significantly different,
resulting in a large spread of $\Delta v$ at each stage of the
journey.  The probe’s maximum velocity is eventually limited by the
minimum value $\delta$ can take (see Discussion).

\subsection{Slingshot Trajectory, Highest $\Delta v$ (maxspeed)}

\begin{figure*}
\begin{center}$
\begin{array}{cc}
\includegraphics[scale = 0.45]{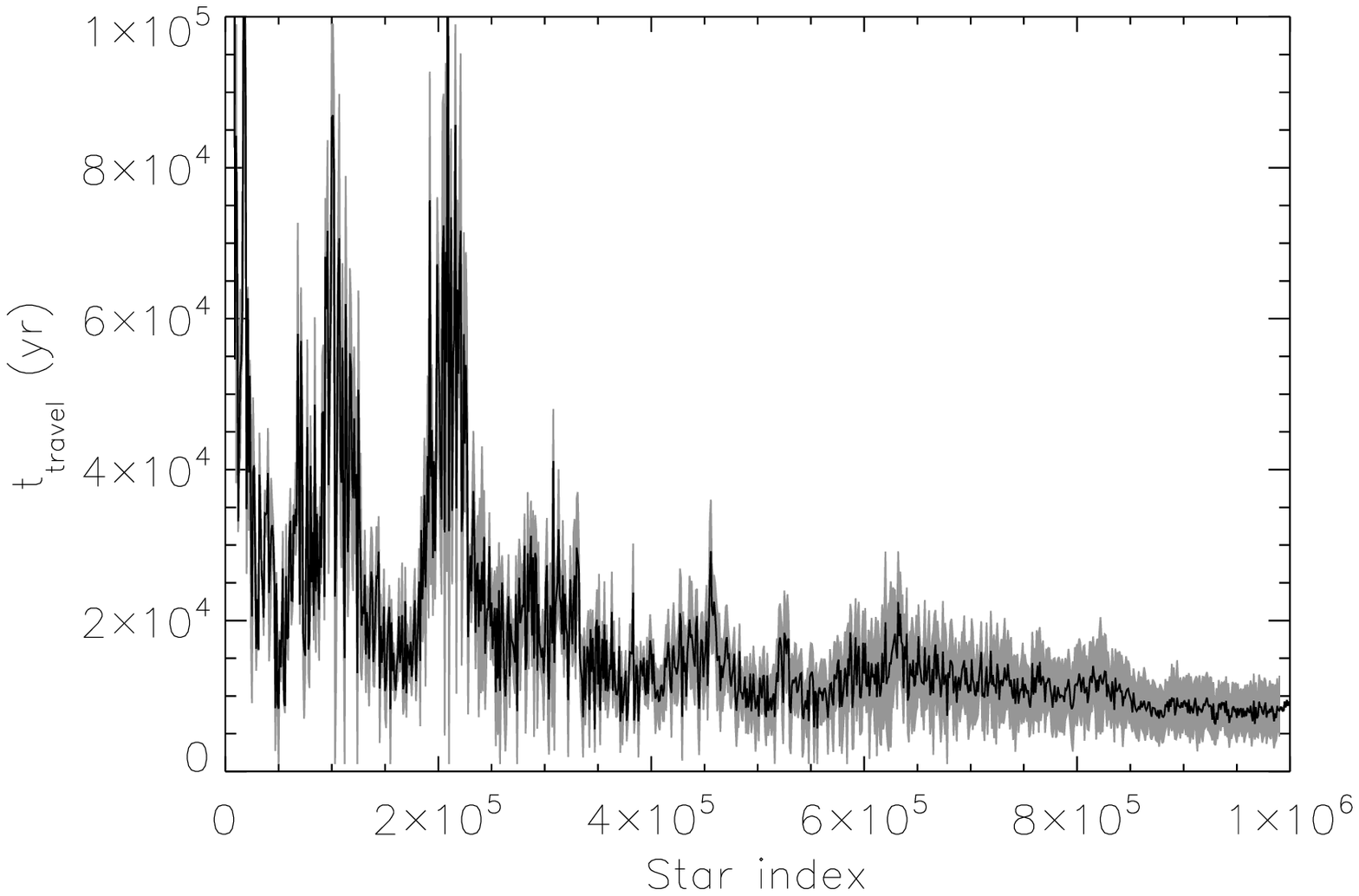} &
\includegraphics[scale = 0.45]{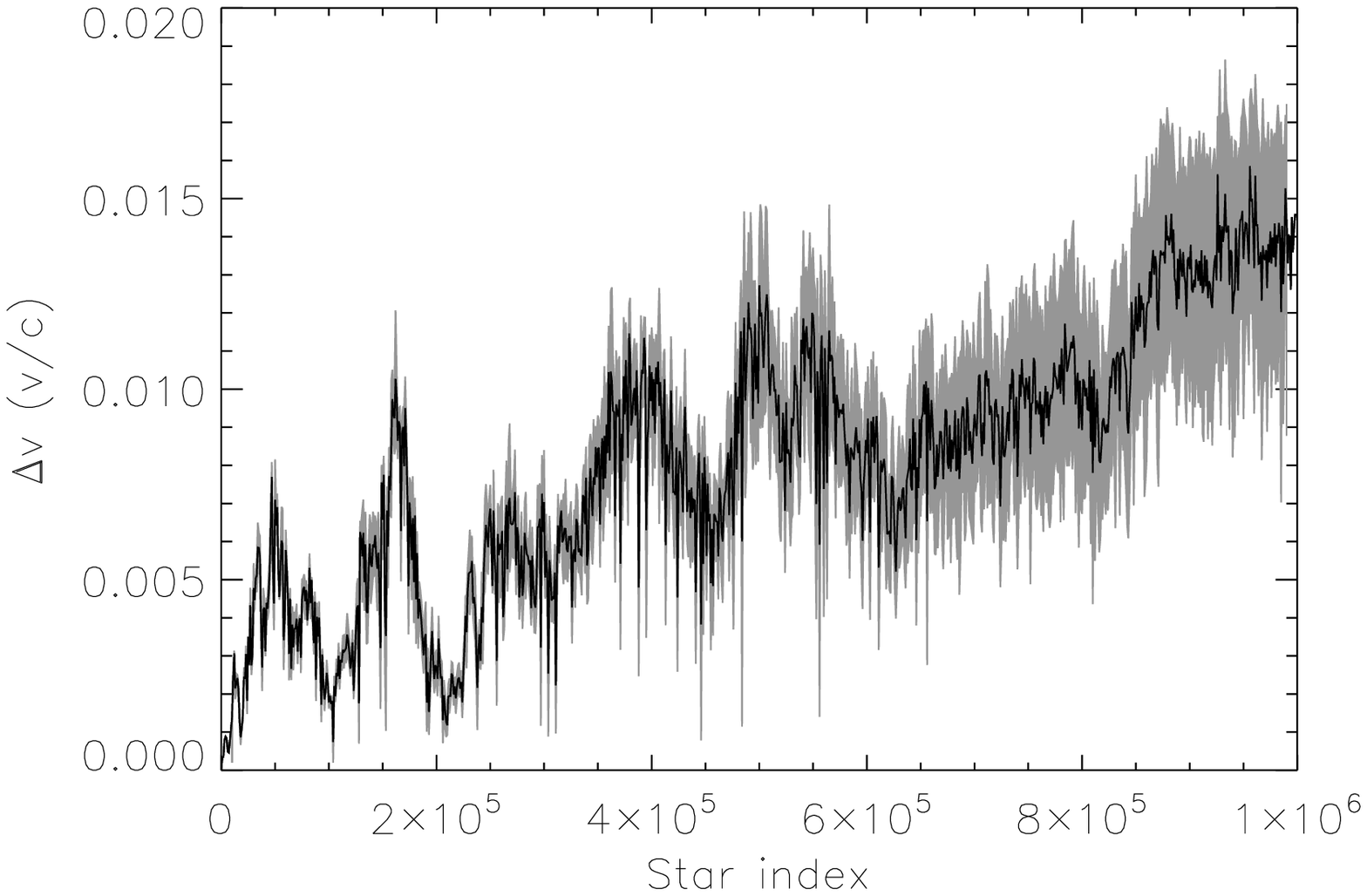} \\
\end{array}$
\caption{\textbf{Left}:The mean travel time between each star for
  probes in the \emph{maxspeed} simulations, averaged over 30
  realisations.  The grey shaded area represents the standard error on
  the mean.  \textbf{Right}: the mean $\Delta v$ generated as a result
  of each slingshot maneouvre, with standard error on the mean shaded
  grey. } \label{fig:proper}
\end{center}
\end{figure*}

\noindent With the probe now attempting to maximise its $\Delta v$, it
pursues a fundamentally different trajectory to the other two cases.
The selection criterion for the next star of the journey is purely to
maximise $\Delta v$ - the distance to the next star is not
considered.  The left hand panel of Figure \ref{fig:proper}
demonstrates the consequences of this.  The travel time indicates
protracted journeys between stars in the early stages of the probe's
trip.  The spikes in travel time are of high significance, and
represent the extent of the simulation box.  No periodic boundary
conditions are applied - as a result, the probe commonly selects
destinations that are at the opposite end of the box, and as a result
must traverse the box's entire length frequently.  The amplitude of
these peaks decreases as the probe begins to boost its speed strongly
(right hand panel of Figure \ref{fig:proper}). By selecting for maximal
$\Delta v$, the probe can achieve velocity increases 2.5 times larger
than the slingshot case.

\section{Discussion}\label{sec:discussion}

%

\subsection{Limitations of the Analysis}

\noindent Before discussing the implications of these results, we
should first note some simplifications made to expedite the analysis.

Probably the most important simplification made was to fix the stars
in position.  While all stars possessed a velocity vector, the
position vectors were never updated.  As the aim was to create a large
number of realisations at modest computational expense, it was felt
that this assumption was reasonable.  However, it does force us to
jettison important dynamical aspects.  For example, in the maxspeed
case, the probe will select destination stars with relative velocity
vectors which are large and negative, i.e the destination star and the
current star are moving towards each other.  This will reduce the travel time
(just as positive relative velocity will increase the travel time).
Future work in this area should investigate this effect.

The probe's motion was also simplified in several ways.  The maximum
$\Delta v$ achievable by the probe is limited by the distance of
closest approach that a probe can make to the star.  From equation
(\ref{eq:maxdelta}), for a star with mass $M_*$ and an effective
radial boundary $R_{eff}$ that the probe cannot cross, there is a
maximum value of rotation angle $\delta$ dependent on the inward
stellarcentric velocity, $u_{i}$ \citep{Gurzadyan1996}:

\begin{equation} \tan \frac{\delta}{2} = \frac{\sqrt{GM_*}}{R_{\rm eff}
    u_{i}} \end{equation}

\noindent Substituting this into equation \ref{eq:deltav}, we obtain:

\begin{equation} \Delta v_{\rm max} =
  \frac{u^2_{esc}}{\frac{u^2_{esc}}{2u_i} + u_i} \end{equation}

\noindent where $u_{esc}$ is the escape velocity at the radial
boundary of the star:

\begin{equation} u_{esc} = \sqrt{\frac{2GM_*}{R_{\rm eff}}}\end{equation}

\noindent The maximum $\Delta v$ achievable becomes quite negligible
as $u_{i}$ exceeds $\sim 0.01 c$ - for example, a star of mass $1
M_{\rm odot}$ and $R_{eff}=R_{\odot}$ gives a maximum $\Delta v$ of
$\sim 10^{-11}c$.  To improve this value, the probe would have to risk
very close approaches with massive, compact objects such as neutron
stars and black holes \citep{Dyson1963}, which could present hazardous
tidal forces upon the craft's hull.

We have ignored relativistic effects in this analysis.  The probe
achieves velocities of $\sim 0.01 c$, which gives a Lorentz factor

\begin{equation} \gamma = \frac{1}{\sqrt{1-
      \left(\frac{v}{c}\right)^2}}=1.00005, \end{equation}

\noindent indicating that classical physics is an acceptable
approximation for this work.  However, future studies that use a
higher initial velocity should be cognisant of this. 
 
\subsection{Implications for Fermi's Paradox}

\noindent Let us now investigate possible constraints on Fermi's
Paradox, and in particular, the Fermi-Hart timescale $t_{FH}$.  If we
assume a constant number density of stars, then we can estimate
$t_{FH}$ from the average total travel times calculated in this study:

\begin{equation} t_{FH} \sim \left(\frac{3 \times 10^{-5}}{\frac{v_{\rm
        initial}}{c}}\right)\left(\frac{N_{star}}{10^6}\right)t_{\rm
    travel}. \label{eq:ourtfh} \end{equation}

\noindent Assuming that the number of stars in the Galaxy, $N_{star} =
10^{11}$, then at $v_{\rm initial} = 3 \times 10^{-5} c$: 

\begin{equation} t_{FH} \sim 10^5 t_{\rm travel} \sim 10^{13} -
  10^{15} \,\mathrm{yr}. \end{equation}

\noindent This confirms that given the low initial velocity we
selected, motivated by current speed records of human-manufactured
probes, we are currently incapable of exploring the Galaxy inside the
Hubble time with a single probe.  However, we should still note that
we are much stricter than we might need to be when imposing the
maximum powered velocity of a spacecraft, and the relative decrease in
travel time depending on probe dynamics is significant (as shown in
Table \ref{tab:params}).  As is well known, if higher initial
velocities are indeed possible, then the Fermi-Hart timescale becomes
amenable.  We can estimate the minimum velocity by rearranging
equation (\ref{eq:ourtfh}) for $v_{\rm initial}/c$:

\begin{equation} \frac{v_{\rm initial}}{c} \sim 3 \times 10^{-5}
  \left(\frac{t_{\rm
      travel}}{t_{FH}}\right)\left(\frac{N_{star}}{10^6}\right) = 3
  \left(\frac{t_{\rm travel}}{t_{FH}}\right),\end{equation}

\noindent i.e. even if $v_{\rm initial} \sim c$, then 

\begin{equation} t_{FH} \sim 3 t_{\rm travel} \sim 10^9 - 10^{11} \,\mathrm{yr} \end{equation}

\noindent which is still quite high (except for possibly the slingshot
case, and even then the velocity boosts possible will be much smaller
than those achieved at low speeds).  From this, we appear to confirm
previous calculations that even when probe dynamics are considered in
more detail, one of two conditions must be satisfied for $t_{FH}$ to
be sufficiently low:

\begin{enumerate}
\item Faster than light travel is possible, or
\item  multiple probes are required  \cite{Tipler1980,Wiley2011}.
\end{enumerate}

\noindent While option 1 has been explored theoretically and found to
be possible for massless particles and for civilisations that can
correctly manipulate space-time (see e.g. \citealt{Crawford1995} for a
review of these concepts), it requires the existence of exotic matter
and energy sources, not to mention staggering technological prowess we
do not yet possess. Option 2 would therefore appear to be the most
plausible choice.  Indeed, probes which carry out a series of stellar flybys
without leaving significant evidence of their passage would not affect
Fermi's Paradox at all, as individual star systems would only be able
to detect these probes briefly (if at all).  It would seem reasonable
then to require that probes leave some sort of beacon or artifact in
their wake as they pass through a system, to signal their existence
and to strengthen Fermi's Paradox.  The manufacture (or replication)
of many beacons becomes an industrial problem on a scale similar to
that of manufacturing multiple probes.

Given our current ability to manufacture large numbers of similar
sized craft for terrestrial uses, it is not unlikely that we can adopt
a similar approach to building probes.  A simple calculation shows
that producing $10^{11}$ Voyager-esque probes would allow humankind to
explore the Galaxy in $10^9$ years.  Given that around $5\times 10^7$
automobiles are produced each year
globally\footnote{http://www.oica.net/}, it seems reasonable to expect
a coordinated global effort could produce the requisite probes within
a few thousand years.  If the probes are made to be self-replicating,
using materials en route to synthesise copies, the exponential nature
of this process cuts down exploration time dramatically
\citep{Wiley2011}.

Whether ETIs create a static fleet of probes launched from one source,
or an exponentially growing fleet of probes replicated in transit from
raw material orbiting destination stars, we argue that these measures
strengthen the Fermi Paradox further when slingshot dynamics are
included, but the relative strength of dynamics versus numbers are
currently unclear.  To answer this, we are repeating the analysis made in
this paper with self-replicating probes to investigate whether
optimised slingshot dynamics are even worthwhile in a cost-benefit
scenario (Nicholson et al, in prep).

Having said this, probes carrying out slingshot trajectories in a
larger, more realistic domain will be able to produce more realistic
trajectories.  This is another important avenue for further
investigation.

\section{Conclusions}\label{sec:conclusions}

\noindent We carried out Monte Carlo Realisation (MCR) simulations of
a single probe traversing a section of the Galaxy, exploring the
effect of slingshot dynamics on the total travel time.  Three
scenarios were explored: the standard scenario where probes travel
under powered flight to their nearest neighbour (powered); traversing
the same path, but utilising slingshot trajectories to boost the
probe's velocity (slingshot); and a third scenario where the probe
selects its next destination in order to maximise the probe's velocity
boost (maxspeed).

We find that allowing the probe to make slingshots can reduce the
probe's total travel time by two orders of magnitude.  The velocity
boost is typically additive, and as a result this velocity boost could
presumably be marginally increased by allowing the probe to explore a
larger domain and a higher quantity of stars.

The speed of the probe was selected to represent current human ability
in unmanned spaceflight.  The exploration time of probes moving at
this speed is sufficiently large that humans would struggle to explore
the Galaxy inside one Hubble time without mass production of probes.
Our work shows that even when optimising probe trajectories are taken into
account, the travel time remains quite large.

\section{Acknowledgements}

DF gratefully acknowledges support from STFC grant ST/J001422/1.  SP gratefully
acknowledges support from a Robert Cormack Bequest Vacation
Scholarship (provided by the Royal Society of Edinburgh).

\bibliographystyle{mn2e} 
\bibliography{probe_staticbox}

\end{document}